\documentstyle[11pt,amsfonts]{article}

\newcommand{\be}{\begin{equation}}
\newcommand{\ee}{\end{equation}}
\newcommand{\bea}{\begin{array}}
\newcommand{\ea}{\end{array}}
\newcommand{\beqa}{\begin{eqnarray}}
\newcommand{\eeqa}{\end{eqnarray}}
\newcommand{\bean}{\begin{eqnarray*}}
\newcommand{\eean}{\end{eqnarray*}}

\def\up#1{\leavevmode \raise.16ex\hbox{#1}}

\newcommand{\gapproxeq}{\lower
 .7ex\hbox{$\;\stackrel{\textstyle >}{\sim}\;$}}
\newcommand{\lapproxeq}{\lower .7ex\hbox{$\;\stackrel
{\textstyle <}{\sim}\;$}}

\newcounter{appendice}

\def\thebibliography#1{{\bf REFERENCES\markboth
 {REFERENCES}{REFERENCES}}\list
 {[\arabic{enumi}]}{\settowidth\labelwidth{[#1]}\leftmargin\labelwidth
 \advance\leftmargin\labelsep
 \usecounter{enumi}}
 \def\newblock{\hskip .11em plus .33em minus -.07em}
 \sloppy
 \sfcode`\.=1000\relax}

\def\BI{{\rm 1\!l}}

\begin{document}

\vskip 1cm
\centerline{ \LARGE Supersymmetric Extension of the Snyder Algebra }

\vskip 2cm

\centerline{{\sc L. Gouba}${}^*$  and    {\sc A. Stern }${}^{**}$ }

\vskip 5mm

\centerline{
*The Abdus Salam International Centre for Theoretical Physics(ICTP)}\centerline{ Strada Costiera 11, 34014, Trieste, 
Italy. $\;\;$ Email: lgouba@ictp.it}
\vskip 5mm

\centerline{**  Dept. of Physics and Astronomy, Univ. of Alabama,}\centerline{
Tuscaloosa, Al 35487, U.S.A.$\;\;$ Email: astern@bama.ua.edu}
\vskip 1cm

\vspace*{10mm}

\normalsize
\centerline{\bf ABSTRACT}

We obtain a minimal supersymmetric extension of the Snyder algebra and study its representations.   The construction differs from the general approach given in Hatsuda and Siegel ({\tt hep-th/0311002}) 
and does not utilize super-de Sitter groups.
The  spectra of the position operators are discrete, implying a lattice description
of space, and the lattice  is compatible with supersymmetry transformations.

\newpage

\vskip 4 cm 
\section{Introduction} 

Snyder showed long ago that the continuous symmetries of space-time can be made consistent with  a  lattice through the construction of a covariant noncommutative algebra.\cite{Snyder:1946qz}
The  algebra that Snyder proposed is a  nontrivial unification of space-time with the Poincar\'e algebra.  The lattice is a result of the discrete representations of Snyder's algebra.  It is not a classical lattice because only one spatial coordinate can be determined in a measurement.  
Here we generalize the Snyder algebra to get a  nontrivial unification of space-time with the super-Poincar\'e algebra.  Our supersymmetric extension is minimal and also has discrete representations corresponding to a spatial lattice.  The lattice  is compatible with supersymmetry transformations, and the lattice spacing is half that of the bosonic case.  The supersymmetry generators act in a nonstandard way on the space-time.  As discussed in \cite{us}, the translation group generated by the momenta is not associated with
 discrete translations on the spatial lattice.
The system presented here might be useful in formulating an alternative   discretization of supersymmetric field theories, where  supersymmetry transformations are consistently implemented on the lattice.\cite{Joseph:2011xy}
Supersymmetry has been shown  useful in improving renormalizability properties of noncommutative field theories,\cite{Ruiz:2000hu} and this may turn out to be the case for the Snyder model as well.    We shall not examine field theory here, but    note that  the lattice appears to be the appropriate setting for studying field theory in  this case.  

Supersymmetric extensions of the Snyder algebra in  arbitrary space-time dimensions were previously constructed  by  Hatsuda and Siegel.\cite{Hatsuda:2003wt}
 Their general approach is based on supersymmetric de Sitter algebras.\cite{Gates:1983nr}
  We shall give an alternative construction which attaches a pair of Grassmann odd spinors to de Sitter space, but does not require  the  spinors to  generate a supersymmetric de Sitter algebra. Nevertheless,  the  super-Poincar\'e  algebra is recovered upon projecting to Minkowski space-time.   The algebra  combines  $N=1$ supersymmetry with  the Snyder algebra, and in contrast to \cite{Hatsuda:2003wt}, the fermionic coordinates are all anticommuting.  

The Snyder algebra and its supersymmetric extensions are characterized by a deformation parameter $\Lambda$, which is proportional to one over the lattice spacing. One can define the action of the supersymmetry generators on superspace  in the $\Lambda\rightarrow\infty$ limit.     Superspace cannot be  defined  for finite $\Lambda$ (except in one-dimension), since the operators associated with the space-time  coordinates  are not simultaneously diagonalizable.  On the other hand, the momentum operators commute, and so we can define a `super-momentum space'  for finite $\Lambda$.
We shall introduce fields on this space and write down  differential representations for the supersymmetry generators.

Discrete representations of the Snyder algebra were examined in \cite{us}, where it was argued that there are two distinct Hilbert spaces. (This is a result of the projection from de Sitter space.) The two are distinguished by $SU(2)$ quantum numbers, which are integer in one case and half-integer in the other.  In our supersymmetric extension of the model we can construct fermionic  raising and lowering operators that combine the integer and half-integer states in a single  graded space.  The raising and lowering operators simultaneously change the spin by $\pm 1/2$, and the location on the lattice to the nearest neighbor.  Care must be taken in defining an involution for the fermionic operators.  Negative norm states result if complex conjugation connects Lorentz spinors in the standard way.  On the other hand,  negative norm states are absent upon adopting an alternative involution which maps one  de Sitter spinor to the other.

The outline of this article is as follows: In section 2 we review the derivation of the Snyder algebra starting from de Sitter space.  De Sitter  spinors are introduced in section 3.  Their projection to Minkowski space yields the minimal supersymmetric Snyder algebra.  Discrete representations are examined in section 4 and an involution is introduced in section 5 which  eliminates negative norm states.  A differential representation for the supersymmetry generators on super-momentum space is given in section 6.  Concluding remarks are made in section 7, including the construction of the extended supersymmetric Snyder algebra.

\section{Snyder algebra}

\setcounter{equation}{0}
The original derivation of the Snyder algebra starts with a de Sitter manifold.  Say that the latter is coordinatized by  $ P_M$, $\;M=0,1,2,3,4$, which are all commuting and subject to the constraint  
\be  P_M P_N\;\eta_{\mbox{\tiny dS}}^{MN} =1\;,\label{dstrspc}\ee
where   $\eta_{\mbox{\tiny dS}}$ is the de Sitter metric  $\eta_{\mbox{\tiny dS}}=$diag$(-1,1,1,1,1)$. We take $ P_M$ to be dimensionless.  Denote by $\ell^{MN}=-\ell^{NM}$ the generators of the de Sitter group with commutation relations
\be [\ell^{MN},\ell^{RS}]= i(\eta_{\mbox{\tiny dS}}^{RM} \ell^{NS}-\eta_{\mbox{\tiny dS}}^{RN} \ell^{MS}+\eta^{SM}_{\mbox{\tiny dS}}\ell^{RN}-\eta_{\mbox{\tiny dS}}^{SN}\ell^{RM})\;,\label{lrntzalgbr}
\ee
Assuming 
 $ P_M$  transforms as a five-vector with respect to the de Sitter group, one has
 \beqa [\ell^{MN}, P^R]&=& i(\eta_{\mbox{\tiny dS}}^{RM}  P^N -\eta_{\mbox{\tiny dS}}^{RN}  P^M)\label{capis5vctr}
\eeqa
The projection of this system to four dimensions was obtained by defining the   four-momentum vector $p^m$, $\;m=0,1,2,3$, according to
\be p^m=\Lambda\frac {P^m}{P^4}  \;,\label{rdctnofmom}\ee and identifying $\ell^{4m}$ with the space-time four vector $x^m$, up to the dimensionful parameter $\Lambda$.  For this Snyder could consistently set
\be  \ell^{mn}=x^mp^n-x^np^m \qquad\qquad \ell^{4m} =\Lambda x^m \label{orbtldsgn} \ee
and then derive the algebra for the two Lorentz vectors from (\ref{lrntzalgbr}) and (\ref{capis5vctr}),
\be [x^m,x^n]=\frac i{\Lambda^2}\ell^{mn} \qquad\quad  [x^m,p^n]=i\Bigl(\eta^{mn} +\frac{p^mp^n}{\Lambda^2}\Bigr)\qquad\quad  [p^m,p^n]=0
\label{xmxn}\;,\ee
where $\eta=$diag$(-1,1,1,1)$ is the Minkowski metric.  $\Lambda$ is a deformation parameter, and the Heisenberg algebra is recovered in the limit $\Lambda\rightarrow\infty$.  From (\ref{dstrspc}) and (\ref{rdctnofmom}) one gets a mass upper bound, $-p^mp_m\le\Lambda^2$.  The spatial coordinates $x_{\tt i}$,  along with the orbital angular momentum $\ell^{\tt ij}$, ${\tt i,j,...}=1,2,3$,  generate the $SO(4)$ subgroup of the de Sitter group.  The discrete spectra of the position operators follows from the discrete representations of $SO(4)$.

\section{Minimal supersymmetric extension}

\setcounter{equation}{0}
To get the supersymmetric version of the algebra we attach two spinor degrees of freedom to de Sitter space.  Define $\Theta_{\tt A}$ and $\bar\Theta_{\tt A}$,  ${\tt A}=1,2,3,4, $ to be  two conjugate four-dimensional spinors which commute with $P^M$ and $\ell^{MN}$, and satisfy the anti-commutation relations
 \be \{\Theta_{\tt A},\bar\Theta_{\tt B}\}=-i\delta_{\tt AB} \qquad\qquad  \{\Theta_{\tt A},\Theta_{\tt B}\}= \{\bar\Theta_{\tt A},\bar\Theta_{\tt B}\}=0 \label{Capthtbrkts} \ee   Assume they give a spin contribution to the generators of the de Sitter group.  Denote by  $j^{MN}=-j^{NM}$ the sum of the orbital and spin contributions,
\be j^{MN}=\ell^{MN} - \bar\Theta\Sigma^{MN}\Theta \;,\label{dstrgnrtrs}\ee
where  $\Sigma^{MN}=-\Sigma^{NM}$ define  the four-by-four spinor representation of the algebra 
\be [\Sigma^{MN},\Sigma^{RS}]= \eta_{\mbox{\tiny dS}}^{RM} \Sigma^{NS}-\eta_{\mbox{\tiny dS}}^{RN} \Sigma^{MS}+\eta^{SM}_{\mbox{\tiny dS}} \Sigma^{RN}-\eta_{\mbox{\tiny dS}}^{SN} \Sigma^{RM}
\ee
Then $\Theta$ and $\bar\Theta $ transform as de Sitter spinors, while $P^M$ remains a five-vector under de Sitter transformations, now generated by $j^{MN}$, 
\beqa [j^{MN},\bar\Theta]&=&i\bar\Theta \Sigma^{MN} \cr &&\cr [j^{MN}, \Theta]&=&-i\Sigma^{MN}\Theta\cr &&\cr [j^{MN}, P^R]&=& i(\eta_{\mbox{\tiny dS}}^{RM}  P^N -\eta_{\mbox{\tiny dS}}^{RN}  P^M)\cr &&\cr [j^{MN},j^{RS}]&=& i(\eta_{\mbox{\tiny dS}}^{RM} j^{NS}-\eta_{\mbox{\tiny dS}}^{RN} j^{MS}+\eta^{SM}_{\mbox{\tiny dS}} j^{RN}-\eta_{\mbox{\tiny dS}}^{SN} j^{RM})\;\label{ttllrntzalgbr}
\eeqa 

To project to four space-time dimensions we again assume
(\ref{rdctnofmom}) and (\ref{orbtldsgn}), leading to the Snyder algebra  (\ref{xmxn}) for the bosonic operators $x^m$ and $p^m$.  For the spinor representations, we choose
\be \Sigma^{mn}=\pmatrix{\sigma^{mn} &\cr & \bar\sigma^{mn}\cr}\qquad\qquad \Sigma^{4m}=\frac 12 \pmatrix{&\sigma^m\cr\bar\sigma^m &\cr}\;,
\ee 
where we follow conventions in Wess and Bagger:\cite{WB}
 $\sigma^0=\bar\sigma^0=-\BI_{2\times 2}$ and  
 $\bar\sigma^{\tt i}=-\sigma^{\tt i}$,   where $\sigma^{\tt i},\;{\tt i}=1,2,3$,  are the three Pauli matrices.  The $2\times 2$ Lorentz matrices $\sigma^{mn}$ and $\bar\sigma^{mn}$ are defined by
\beqa \sigma^{mn} =\frac 14 (\sigma^m\bar \sigma^n - \sigma^n\bar \sigma^m) &\qquad\quad &\bar \sigma^{mn} =\frac 14 (\bar\sigma^m \sigma^n -\bar \sigma^n \sigma^m)\label{lrntzspin} \eeqa 
They satisfy the identities
 \beqa  2\sigma^{mn}\sigma^s &=& \eta^{ms}\sigma^n
-\eta^{ns}\sigma^m +i\epsilon^{mnsr}\sigma_r \cr & &\cr   2\sigma^s\bar \sigma^{mn} &=& -\eta^{ms}\sigma^n
+\eta^{ns}\sigma^m +i\epsilon^{mnsr}\sigma_r  \cr & &\cr 2\bar\sigma^{mn}\bar\sigma^s &=& \eta^{ms}\bar\sigma^n
-\eta^{ns}\bar\sigma^m -i\epsilon^{mnsr}\bar\sigma_r  \cr & &\cr   2\bar\sigma^s\sigma^{mn} &=& -\eta^{ms}\bar\sigma^n
+\eta^{ns}\bar\sigma^m -i\epsilon^{mnsr}\bar\sigma_r \cr &&\cr
 [\sigma^{mn},\sigma^{rs}]&=&\eta^{rm} \sigma^{ns}-\eta^{rn} \sigma^{ms}+\eta^{sm} \sigma^{rn}-\eta^{sn} \sigma^{rm}\cr &&\cr
 [\bar\sigma^{mn},\bar\sigma^{rs}]&=&\eta^{rm}\bar \sigma^{ns}-\eta^{rn} \bar\sigma^{ms}+\eta^{sm}\bar \sigma^{rn}-\eta^{sn} \bar\sigma^{nm}
\eeqa

Next we express  the   two  four-dimensional spinors
$\Theta$ and $\bar\Theta$ in terms of four two-dimensional Lorentz spinors   $Q_\alpha $, $\bar Q_{\dot \alpha}$,  $\theta^\alpha $ and $\bar \theta^{\dot \alpha}$, $\;
\alpha,\dot\alpha=1,2$.  We write $\bar\Theta$ as a row matrix and $\Theta$ as a column matrix according to
\be \bar \Theta=\frac 1{\sqrt{\Lambda}}\pmatrix{\Lambda\theta \;&\;\bar Q +i\theta\sigma^n p_n }\qquad\qquad \Theta=\frac 1{\sqrt{\Lambda}}\pmatrix{-Q+ip_n\sigma^n\bar\theta\cr\Lambda\bar\theta }\label{redcaptht}\;,\ee $\Lambda$ being the same dimensionful parameter appearing in (\ref{rdctnofmom}).
Since $\Theta$ and $\bar\Theta$ commute with $p^m$, the   four two-dimensional spinors must also commute with $p^m$. 
From the anti-commutation relations (\ref{Capthtbrkts}) for $\Theta$ and $\bar\Theta$, it follows: 

\noindent $i)$ that $\theta^\alpha $ and $\bar \theta^{\dot \alpha}$ are all anti-commuting Grassmann coordinates\footnote{This is  in contrast to \cite{Hatsuda:2003wt}, where the anticommutators of fermionic coordinates are in general  a linear combination of space-time coordinates, Lorentz generators and additional $SO(N)$ generators.}, i.e.,
\be  \{\theta^\alpha,\theta^\beta\}=\{\bar \theta^{\dot \alpha},\bar \theta^{\dot \beta}\} =\{\theta^\alpha,\bar \theta^{\dot \beta}\}=0\;,\label{grsmncrdnts}
\ee
$ii)$ that $Q_\alpha $ and $\bar Q_{\dot \alpha}$ are canonically  conjugate to  $\theta^\alpha $ and $\bar \theta^{\dot \alpha}$, respectively, 
  \be
 \{Q_\alpha,\theta^\beta\}=i\delta_{\alpha}^{\beta}\qquad \quad \{\bar Q_{\dot \alpha},\bar\theta^{\dot\beta}\}=-i\delta_{\dot \alpha}^{\dot \beta} \qquad\quad \{Q_\alpha,\bar\theta^{\dot\beta}\}= \{\bar Q_{\dot \alpha},\theta^\beta\}=0 \label{susyhisnbrg}\;,\ee
and $iii)$ that $Q_\alpha$ and $\bar Q_{\dot \alpha}$  are  super-translation generators,
\be\{Q_\alpha,\bar Q_{\dot \alpha}\}= 2{\sigma_{\alpha\dot\alpha}}^m p_m\qquad\quad\{Q_\alpha,Q_\beta\}=\{\bar Q_{\dot \alpha},\bar Q_{\dot \beta}\}=0
\label{susyalgbr}\ee
Finally, since  $\Theta$ and $\bar\Theta$ commute with  coordinates $x^m$, it follows:

\noindent   $iv)$ that $\theta$ and $\bar\theta$ also commute with $x^m$, while
\beqa [Q_\alpha,x^m]=\Bigl(  \sigma^m + \frac {\sigma^np_np^m}{\Lambda^2}\Bigr)_{\alpha\dot\alpha}\bar\theta^{\dot\alpha} &\qquad\quad &  [\bar Q_{\dot \alpha},x^m]= -\theta^\alpha\Bigl(  \sigma^m + \frac {\sigma^np_np^m}{\Lambda^2}\Bigr)_{\alpha\dot\alpha}\quad \label{4dsnyderQQdx}\eeqa

$Q_\alpha$, $\bar Q_{\dot  \alpha}$ and $p^m$ generate the  $N=1$ super-translation group.   Upon including the Lorentz generators $j^{mn}$, which using (\ref{redcaptht}) can be expressed as \be  j^{mn}=x^mp^n-x^np^m  +\theta\sigma^{mn}Q-\bar Q \bar\sigma^{mn}\bar\theta -\epsilon^{mnrs}(\theta\sigma_r\bar\theta) p_s \;, \label{Lrntzgnrtrs}\ee
we get the super-Poincar\'e group.  From the commutation relations with $j^{mn}$,
\beqa  [Q_\alpha,j^{mn}]= i(\sigma^{mn}Q)_\alpha &\qquad\quad &  [\bar Q_{\dot \alpha},j^{mn}]= -i(\bar Q \bar \sigma^{mn})_{\dot \alpha}\cr &&\cr
[\theta^\alpha,j^{mn}]= -i(\theta\sigma^{mn})^\alpha  &\qquad\quad &  [\bar \theta^{\dot \alpha},j^{mn}]=i(\bar\sigma^{mn}\bar\theta)^{\dot\alpha}\cr &&\cr
[x^r,j^{mn}]= i(x^m\eta^{rn}- x^n\eta^{rm})  &\qquad\quad &  [p^r,j^{mn}]= i(p^m\eta^{rn}- p^n\eta^{rm})  \label{spnrvctrtrns}\;,\eeqa
it follows that $Q_\alpha$, $\bar Q_{\dot \alpha}$,
$\theta^\alpha $ and $\bar \theta^{\dot \alpha}$ transform as Lorentz spinors, while $x^m$ and $p^m$ as Lorentz vectors.

 Eqs. (\ref{grsmncrdnts})-(\ref{4dsnyderQQdx}), along with (\ref{xmxn}), defines a minimal supersymmetric extension of the Snyder algebra.  It is expressed above in terms of  $\theta^\alpha$,  $\bar\theta^{\dot\alpha}$ and $x^m$, in addition to the $N=1$ super-translation generators $Q_\alpha$, $\bar Q_{\dot  \alpha}$ and $p^m$.   Alternatively, instead of  $x^m$, we can define the  following space-time coordinates in the supersymmetric theory:
\beqa   X^m=\frac 1{\Lambda}j^{4m} &=&x^m+\frac 1{2\Lambda^2}\Bigl(\bar Q\bar\sigma^m Q+i\theta\sigma^np_n\bar\sigma^mQ-i\bar Q\bar\sigma^m\sigma^np_n\bar \theta\Bigr)\cr&&\cr&&\qquad\;\;-\frac 1{\Lambda^2}  p^m\theta\sigma^np_n\bar\theta-\frac 12 \Bigl(1-\frac{p^np_n}{\Lambda^2} \Bigr)\theta\sigma^m\bar\theta\;\label{capx}\eeqa 
It forms a closed algebra with the Poincar\'e generators.  In contrast to (\ref{xmxn}), one gets
\be [X^m,X^n]=\frac i{\Lambda^2}j^{mn} \qquad\quad  [X^m,p^n]=i\Bigl(\eta^{mn} +\frac{p^mp^n}{\Lambda^2}\Bigr)\qquad\quad  [p^m,p^n]=0
\label{capxmxn}\;,\ee where $j^{mn}$ are the Lorentz generators (\ref{Lrntzgnrtrs}).    The latter can be re-expressed in terms of $X^n$ using  (\ref{capx}).  $X^n$'s commutators with the spinors are given by
\beqa [X^m,\theta]&=&\frac  i{2\Lambda^2}\;(\bar Q+i  \theta \sigma^n p_n)\bar \sigma^m \cr &&\cr  [X^m,\bar\theta]&=&\frac  i{2\Lambda^2}\;\bar\sigma^m (Q-ip_n\sigma^n\bar\theta) \cr &&\cr [X^m,\bar Q+i  \theta \sigma^n p_n]&=&\frac  i{2}\theta \sigma^m \cr &&\cr  [X^m, Q-i  \sigma^n p_n\bar \theta]&=&\frac  i{2} \sigma^m\bar \theta\label{cmtrsfcX}
\eeqa
Eqs. (\ref{capxmxn}) and (\ref{cmtrsfcX}), along with  (\ref{grsmncrdnts})-(\ref{susyalgbr}), give an alternative definition of the supersymmetric Snyder algebra.  Due to the nonstandard commutation relations of  $Q_\alpha$, $\bar Q_{\dot\alpha}$ and $p_m$  with the coordinate operators $X^m$ (or $x^m$), the super-translation group acts in a nonstandard fashion on space-time.  

\section{Discrete representations}

\setcounter{equation}{0}
We regard $X_{\tt i}$, ${\tt i}=1,2,3$, as the position operators for the supersymmetric Snyder algebra.  They are the spatial components of $X^n$.  From  (\ref{capxmxn}), $X_{\tt i}$, along with $j^{\tt ij}$,  generate another  $SO(4)$ group.  It follows that $X_{\tt i}$, like $x_{\tt i}$, have  discrete spectra.
Here we can define
\be A_{\tt i} =\frac 12\Bigl( J_{\tt i} +\Lambda X_{\tt i}\Bigr)\qquad\qquad  B_{\tt i} =\frac 12\Bigl( J_{\tt i}  -\Lambda X_{\tt i}\Bigr)\;,\label{su2chargs}\ee  where $ J_{\tt i}=\frac12\epsilon_{\tt ijk}\,j^{\tt jk} $ is the rotation generator.
They satisfy two $su(2)$ algebras
\beqa [ A_{\tt i}, A_{\tt j}] &=&i\epsilon_{\tt ijk} A_{\tt k}\cr & &\cr  [ B_{\tt i},  B_{\tt j}] &=&i\epsilon_{\tt ijk}  B_{\tt k}\cr & &\cr [ A_{\tt i}, B_{\tt j}] &=&0 \; \label{ofralgbr} \eeqa 
$ A_{\tt i} A_{\tt i}$, $ B_{\tt i} B_{\tt i}$, $ A_3$ and  $ B_3$ form a complete set of commuting operators.\footnote{The Casimir operators  $ A_{\tt i} A_{\tt i}$ and $ B_{\tt i} B_{\tt i}$ are independent, unlike their counterparts in the non supersymmetric theory.\cite{us}} We denote their eigenvectors by $|j_A,j_B,m_A,m_B>$, 
\beqa
 A_{\tt i} A_{\tt i}\;|j_A,j_B,m_A,m_B> &=& j_A(j_A+1)\;|j_A,j_B,m_A,m_B>\cr & &\cr
 B_{\tt i} B_{\tt i}\;|j_A,j_B,m_A,m_B> &=& j_B(j_B+1)\;|j_A,j_B,m_A,m_B>\cr & &\cr
 A_3\;|j_A,j_B,m_A,m_B> &=& m_A|j_A,j_B,m_A,m_B>\cr & &\cr
 B_3\;|j_A,j_B,m_A,m_B> &=& m_B\;|j_A,j_B,m_A,m_B> \;,\label{eigneqAB} \eeqa
where $ m_A=-j_A,1-j_A,...,j_A\;,$   and   $ m_B=-j_B,1-j_B,...,j_B\;$.   $j_A$ and $j_B$ take values $0,\frac 12,1,\frac 32,...\;,$ and label the $SO(4)$ representations. 
 $|j_A,j_B,m_A,m_B>$  is also an eigenvector of  $ J_3$ and  $ X_3$,  whose  corresponding eigenvalues are   $m_A+m_B$ and $(m_A-m_B)/{\Lambda}$, respectively.  
Applying  $ A_\pm$ or $ B_\pm$ changes $|J_3|$ by $1$ and $|X_3|$ by $1/\Lambda$. 

 $SO(4)$ group representations are also present in the non-supersymmetric version of the theory,  and there $m_A$ and $m_B$ are either both integer or both half-integer.\cite{us}  It followed that the associated (orbital) angular momentum operators had  integer eigenvalues,  and also that the eigenvalues of the position operators $x_{\tt i}$ were evenly spaced at intervals of  $\Lambda^{-1}$. We show below that for the supersymmetric Snyder algebra,   eigenvalues of the position operators $X_{\tt i}$ are evenly spaced at intervals of  $\frac 12\Lambda^{-1}$,  and that both integer and half-integer values of  $m_A$ and $m_B$  occur in  the  representation of the graded algebra. 
 This means that  both integer and half-integer values  for $j_A$ and   $j_B$ occur in a representation of the supersymmetry algebra.

From (\ref{spnrvctrtrns}) and (\ref{cmtrsfcX})   one can  construct various linear combinations of the spinors which act as raising and lowering operators with respect to the eigenvalues of the  operators $X_{\tt i}$ and  $J_{\tt i}$, or equivalently, $A_{\tt i}$ and  $B_{\tt i}$.  For the case of ${\tt i}=3$, we can define the raising and lowering operators $a^{\tt A}_\pm$, ${\tt A}=1,2,3,4$, according to
\beqa  \sqrt{2}\; a^1_\pm&=&\bar\Theta_1\mp i\bar\Theta_3\;\;=\;\;\sqrt{\Lambda}\theta^1 \mp \frac i{\sqrt{\Lambda}}(\bar Q+i\theta\sigma^np_n)_1\cr &&\cr
 \sqrt{2}\; a^2_\pm&=&\bar\Theta_2\pm i\bar\Theta_4\;\;=\;\;\sqrt{\Lambda}\theta^2 \pm  \frac i{\sqrt{\Lambda}}(\bar Q+i\theta\sigma^np_n)_2\cr &&\cr 
 \sqrt{2}\; a^3_\pm&=&\Theta_3\pm i\Theta_1\;\;=\;\;\sqrt{\Lambda}\bar\theta^1 \mp  \frac i{\sqrt{\Lambda}}( Q-i\sigma^np_n\bar\theta)_1\cr &&\cr 
 \sqrt{2}\; a^4_\pm&=&\Theta_4\mp i\Theta_2\;\;=\;\;\sqrt{\Lambda}\bar\theta^2 \pm  \frac i{\sqrt{\Lambda}}( Q-i\sigma^np_n\bar\theta)_2\eeqa
Their nonvanishing anticommutators are
\be\{a^1_\mp, a^3_{\pm}\}=\{a^2_{\pm},a^4_{\mp}\}=\pm 1\label{aanticmtrs}\ee
From  (\ref{spnrvctrtrns}) and (\ref{cmtrsfcX}) one has \be [X_3, a^{\tt A}_\pm]=\pm \frac 1{2\Lambda}\;a^{\tt A}_\pm\; \label{ralx3} \ee 
\be [J_3, a^{1,4}_\pm]= \frac 12 a^{1,4}_\pm\quad\qquad  [J_3, a^{2,3}_\pm]= -\frac 12 a^{2,3}_\pm\label{J3cmtrwapm}\ee
 Say that    $|m_A,m_B>$  is  an eigenvector of  $ X_3$ and $J_3$ with eigenvalue is $(m_A-m_B)/{\Lambda}$ and $m_A+m_B$, respectively, where for convenience we ignore the dependence on the indices $j_A$ and $j_B$.
Then  $a^{\tt A}_\pm|m_A,m_B>$ are also eigenvectors of  $ X_3$ and $J_3$.  From (\ref{ralx3}), the  $X_3$ eigenvalue  of $a^{\tt A}_\pm|m_A,m_B>$ is $(m_A-m_B\pm\frac 1{2})/{\Lambda}$. 
From (\ref{J3cmtrwapm}), the $ J_3$  eigenvalue of $a^{1,4}_\pm|m_A,m_B>$  is   $m_A+m_B+\frac 1{2}$, while the  $ J_3$  eigenvalue of  $a^{2,3}_\pm|m_A,m_B>$  is  $m_A+m_B-\frac 1{2}$.     So 
application of  $ a^{\tt A}_\pm$  simultaneously changes $|J_3|$ by $\frac12$ and $|X_3|$ by $\frac 1{2\Lambda}$. 
Up to degenerate states,
\beqa a^{1,4}_+|m_A,m_B>&\sim & \Big|m_A+\frac 12,m_B\Big>  \cr&&\cr
 a^{1,4}_-|m_A,m_B>&\sim &\Big |m_A,m_B+\frac 12\Big>\cr&&\cr a^{2,3}_+|m_A,m_B>&\sim & \Big|m_A,m_B-\frac 12\Big>  \cr&&\cr a^{2,3}_-|m_A,m_B>&\sim & \Big|m_A-\frac 12,m_B\Big> \label{actnfcaops}\eeqa
It follows that, unlike what happens in the non-supersymmetric theory, both integer and half-integer values of   $m_A$ and $m_B$ occur in  the representations of the supersymmetric algebra.  Consequently,  both integer and half-integer values of   $j_A$ and $j_B$ occur in  the representations.   In the non-supersymmetric theory,
 the eigenvalues of  the position operator  are regularly spaced at intervals of  $\Lambda^{-1}$, whereas here they  are spaced at intervals of $\frac 12\Lambda^{-1}$.    $a^{\tt A}_\pm$ were defined to raise and lower eigenvalues associated with the $3-$direction.  Raising and lowering operators can also be constructed for the $1-$ and $2-$directions, and they yield the same spectra for the position and angular momentum operators. 

\section{Involutions}

\setcounter{equation}{0}
There are two approaches to  introducing an involution, or complex conjugation, of the algebra.   The first which we discuss below connects Lorentz spinor  $\theta$ with $\bar\theta$ and $Q$ with $\bar Q$ in the usual way.  It leads to negative norm states.  No negative norm states result from an alternative involution.  The latter maps one  de Sitter spinor to the other.

Complex conjugation (which we denote by $*$) in four space-time dimensions standardly relates spinor representations to conjugate representations according to
\be (\theta^\alpha)^*=\bar \theta^{\dot\alpha}\qquad\quad (Q_\alpha)^*=\bar Q_{\dot\alpha}\,
\label{cclrntspnr}\ee
Its action on the de Sitter spinors $\Theta$ and $\bar\Theta$ is then
\be (\bar\Theta^*)=\pmatrix{ &\BI\cr-\BI &\cr}\Theta \qquad\qquad(\Theta^*)=\bar\Theta\pmatrix{ &\BI\cr-\BI &\cr}\label{cconcpth} \ee
It is then easy to show that  
 \be \{\Theta_{\tt A}^*,\bar\Theta_{\tt B}^*\}=i\delta_{\tt AB} \qquad\qquad  \{\Theta_{\tt A}^*,\Theta_{\tt B}^*\}= \{\bar\Theta_{\tt A}^*,\bar\Theta_{\tt B}^*\}=0 \,\label{stranticmtr}\ee 
This means that complex is consistent with the supersymmetric Snyder algebra  generated by $x^m$, $p_m$, $\theta^\alpha$  $,\bar\theta^{\dot\alpha}$, $ Q_\alpha$ and $\bar Q_{\dot\alpha}$, 
 since all anticommutators between Lorentz spinors followed from (\ref{Capthtbrkts}).  We assume that $p_m$ and $x^m$ are real.  It follows that  all of the de Sitter group generators $X^m$ and $j^{mn}$ are also real. 

 For the raising and lowering operators we get
\be (a_\pm^1)^*=a_\mp^3 \qquad\quad (a_\pm^2)^*=a_\mp^4\,\label{ccfapm}
\ee
From the anticommutation relations (\ref{aanticmtrs}), the norm-squared of eigenvectors of $A_3$ and $B_3$ are related by 
\beqa  \Big|a^1_+|m_A,m_B>\Big|^2&=&-\;\Big|a^3_-|m_A,m_B>\Big|^2-\; \Big||m_A,m_B>\Big|^2\cr &&\cr
 \Big|a^2_-|m_A,m_B>\Big|^2&=&-\;\Big|a^4_+|m_A,m_B>\Big|^2-\; \Big||m_A,m_B>\Big|^2\;,\eeqa
implying the existence of negative norm states.
Alternatively, one can define
\be b^{\tt A}_-=\bar\Theta_{\tt A}\qquad \qquad b^{\tt A}_+=i\Theta_{\tt A}\;,\ee which
satisfy the usual algebra of fermionic creation and annihilation operators, 
\be\{b_+^{\tt A},b_-^{\tt B}\}=\delta^{\tt AB} \qquad\quad\{b_+^{\tt A}, b_+^{\tt B}\}=\{b_-^{\tt A}, b_-^{\tt B}\}=0\ee 
However from (\ref{cconcpth}), $b_+^{\tt A}$ is not the complex conjugate of $b_-^{\tt A}$, again implying the existence of negative norm states, e.g.
$\Big|b^1_-|m_A,m_B>\Big|^2=-\Big|b^3_+|m_A,m_B>\Big|^2$.
We note from (\ref{actnfcaops}), that while $b_\pm^{\tt A}$ acting  on  $|m_A,m_B>$  is  an eigenvector of  $ J_3$, it is not an eigenvector of  $ X_3$, $ A_3$ or $B_3$.

On the other hand, negative norm states are absent if we replace the  $*$ by another involution-operation, which we denote by $\star$, satisfying $(x^\star)^\star =x$, 
$(xy)^\star=y^\star x^\star$ and $i^\star=-i$.  We define it
 in terms of the de Sitter spinors according to
\be  \bar\Theta_{\tt A}^\star =i\Theta_{\tt A} \label{stroncpth}\;, \ee in contrast to (\ref{cconcpth}).
(\ref{stranticmtr}) is again satisfied (with $\star$ now replacing $*$), and so the $\star$ involution is  consistent with the supersymmetric Snyder algebra.  We assume that $p_m$ and $x^m$ are real with respect to $\star$. In order that the de Sitter generators $X^m$ and $j^{mn}$ are real under the $\star$ involution we need that
\be  \sigma^{m\star}=-\bar\sigma^m   \qquad\quad \bar\sigma^{m\star}=-\sigma^m \label{strsgma} \ee
The action of the $\star$ involution on  Lorentz spinors is more involved than the previous complex conjugation (\ref{cclrntspnr}).  Using  (\ref{redcaptht}) and (\ref{strsgma}),  we get
\beqa  (\theta^\alpha)^\star &=&-\frac i\Lambda (Q-i \sigma^np_n\bar\theta )_{\alpha}\cr &&\cr(\bar \theta^{\dot\alpha})^\star &=&\frac i\Lambda (\bar Q+i\theta \sigma^np_n )_{\dot\alpha}\cr &&\cr Q_{\alpha}^\star &=& -\frac 1\Lambda ( \bar Q\bar\sigma^n)^\alpha p_n-i\Lambda \theta^\alpha \Bigl(1-\frac{p^np_n}{\Lambda^2}\Bigr)\cr &&\cr \bar Q_{\dot\alpha}^\star &=&-\frac 1\Lambda ( \bar\sigma^nQ)^{\dot\alpha} p_n+ i\Lambda\bar\theta^{\dot\alpha}\Bigl(1-\frac{p^np_n}{\Lambda^2}\Bigr)\eeqa  

From (\ref{stroncpth}), the $\star$ involution of the raising and lowering operators is given by
\be  (a_\pm^{1})^\star=\mp a_\mp^{3}\qquad\quad (a^2_\pm)^\star=\pm a_\mp^{4}\;,\label{cvlnapm}\ee in contrast to (\ref{ccfapm}), or simply, 
\be   (b_\pm^{\tt A})^\star=b_\mp^{\tt A}\ee
We can then introduce a set of states 
  $\{|cv> \}$,  corresponding to   Clifford vacuum states  , i.e.,
$ b_-^{\tt A}|cv>=0\;.$   They should form a representation of the (non-supersymmetric)  Snyder algebra, i.e.,  the algebra generated by the bosonic operators $x_{\tt i}$  and $p_{\tt i}$, since these operators commute with $b_\pm^{\tt A}$.  Two such infinite-dimensional representations were found in \cite{us}.   To the Clifford vacuum  we add all states obtained by acting with $b_+^{\tt A}$ to  obtain a representation  of the supersymmetric theory.  There are no negative norm states in this case, because hermitian conjugation is with respect to $\star$, i.e., hermiticity for any vectors $|\psi>$ and $|\phi >$ here means $<\psi|\phi>^\star=<\phi|\psi>$.

\section{Momentum-dependent superfields}

\setcounter{equation}{0}
Superspace is standardly coordinatized by  space-time coordinates and Grassmann odd variables.    This is not possible for finite $\Lambda$, since the space-time coordinates (either $x^m$ or $X^m$) do not commute amongst themselves in this case, and moreover, they have  discrete spectra.  On the other hand,  since  $[p_m,p_n]=0$, we can define a momentum superspace spanned by $p^m$, $\theta^\alpha $ and $\bar \theta^{\dot \alpha}$, and then write down fields on this space.   Using  $\frac\partial{\partial\theta^\alpha}\theta^\beta=\delta^\beta_\alpha$, $\frac\partial{\partial\bar\theta^{\dot\alpha}}\bar\theta^{\dot\beta}=\delta^{\dot\beta}_{\dot \alpha}$ and $\frac\partial{\partial\theta^{\alpha}}\bar\theta^{\dot\beta}=\frac\partial{\partial\bar\theta^{\dot\alpha}}\theta^{\beta}=0$,  
one can represent the four spinors in (\ref{redcaptht}) by
\be \bar \Theta=\pmatrix{\sqrt{\Lambda}\theta \;&\;-\frac i{\sqrt{\Lambda}}\frac\partial{\partial\bar\theta} }\qquad\qquad \Theta=\pmatrix{-\frac i{\sqrt{\Lambda}}\frac\partial{\partial\theta}\cr\sqrt{\Lambda}\bar\theta }\ee 
Then
 $Q_\alpha$, $\bar Q_{\dot \alpha}$ , $x^m$  and $j^{mn}$ are given by  differential operators  
\beqa
 Q_\alpha &=&i\Bigl(\frac\partial{\partial\theta^\alpha}
+{\sigma_{\alpha\dot\alpha}}^mp_m\bar\theta^{\dot\alpha}\Bigr)\cr & &\cr \bar Q_{\dot \alpha} &=&-i\Bigl(\frac\partial{\partial\bar\theta^{\dot \alpha}}+\theta^\alpha {\sigma_{\alpha\dot\alpha}}^mp_m\Bigr) \cr & &\cr X^m&=&i \Bigl(\frac\partial{\partial p_m} +\frac{p^mp^n}{\Lambda^2}\frac\partial{\partial p^n}\Bigr)-\frac 12\theta\sigma^m\bar\theta+\frac1{2\Lambda^2}
\frac\partial{\partial\bar\theta}\bar\sigma^m
\frac\partial{\partial\theta} \cr & &\cr j^{mn}&=&i\Bigl(-p_m\frac\partial{\partial p_n}+p_n\frac\partial{\partial p_m}+\theta\sigma^{mn}\frac\partial{\partial\theta}
+\frac\partial{\partial\bar\theta}\bar\sigma^{mn}
\bar\theta\Bigr)\label{sprsndrdfrntnlrep}\eeqa

As is usual,  one can construct  fermionic operators $D_\alpha$ and $\bar D_{\dot \alpha}$ which anticommute with the super-translation generators $Q_\alpha$ and $\bar Q_{\dot\alpha}$, and use them   to reduce  supersymmetric representations.  They are
\beqa
 D_\alpha &=&i\Bigl(\frac\partial{\partial\theta^\alpha}
-{\sigma_{\alpha\dot\alpha}}^mp_m\bar\theta^{\dot\alpha}\Bigr)\cr & &\cr \bar D_{\dot \alpha} &=&-i\Bigl(\frac\partial{\partial\bar\theta^{\dot \alpha}}-\theta^\alpha {\sigma_{\alpha\dot\alpha}}^mp_m\Bigr)
\;,\eeqa satisfying
\be\{D_\alpha,\bar D_{\dot \alpha}\}= -2{\sigma_{\alpha\dot\alpha}}^m p_m\qquad\quad\{D_\alpha,D_\beta\}=\{\bar D_{\dot \alpha},\bar D_{\dot \beta}\}=0\label{DDdgralgbr}\ee
A chiral field  
 $ \Phi$ on   super-momentum space  satisfies 
\be \bar D_{\dot\alpha}  \Phi=0\; \ee
This is solved by \be \Phi= {\cal F}(p,\theta)\;e^{-\theta p\cdot \sigma\bar\theta} \ee
Assuming ${\cal F}(p,\theta)$ to be a bosonic field, it can be expanded in terms of two bosonic component fields and a spinor femionic field. We define the action of an operator ${\cal O}$ on the function ${\cal F}(p,\theta)$ according to 
$ {\cal O} \Phi= [{\cal O}{\cal F}]\;e^{-\theta p\cdot \sigma\bar\theta}$.  Applying the super-translation generators, one gets
\beqa   Q_\alpha {\cal F} &=& i \frac {\partial{\cal F}}{\partial \theta^\alpha}  \cr &&\cr\bar  Q_{\dot\alpha} {\cal F} &=& -2i(\theta \sigma\cdot p)_{\dot\alpha}{\cal F}  \label{qqbroncrl}\eeqa 

Similar considerations can be made for  anti-chiral fields $\bar{\Phi}$, which satisfy
\be D_\alpha  \bar\Phi=0\; \ee It is solved by \be \bar{ \Phi}= \bar {\cal F}(p,\bar\theta)\;e^{\theta p\cdot \sigma\bar\theta} \ee
If we define the action of an operator ${\cal O}$ on the function $\bar {\cal F}(p,\bar\theta)$ by $ {\cal O}\bar \Phi= [{\cal O}\bar {\cal F}]\;e^{\theta p\cdot\sigma\bar\theta}$, then in contrast (\ref{qqbroncrl}),
\beqa   Q_\alpha \bar{\cal F} &=& 2i(p\cdot \sigma\bar\theta)_\alpha \bar{\cal F} \cr &&\cr \bar Q_{\dot\alpha}\bar {\cal F} &=&- i \frac{\partial \bar {\cal F}}{\partial\bar \theta^{\dot\alpha} } \eeqa 

\section{Concluding remarks}
\setcounter{equation}{0}

Our derivation of the supersymmetric Snyder algebra closely follows Snyder's work.\cite{Snyder:1946qz}  In particular, it uses his projection  to Minkowski space-time.  Other projections from de Sitter space have been considered.\cite{Hatsuda:2003wt}
 Snyder's algebra also has  been derived  starting from  relativistic  particle dynamics.\cite{Jaroszkiewicz}, \cite{Romero:2004er},\cite{Romero:2006pe},\cite{Banerjee:2006wf},\cite{Chatterjee:2008bp},\cite{Stern:2010ri}  Particle dynamics on Snyder space has also been studied in \cite{Mignemi:2011gr},\cite{Lu:2011fh}.
 Ref. \cite{Stern:2010ri}, in particular,  begins from the reparametrization invariant  action of a relativistic particle.  One arrives at the Snyder algebra (or more precisely, its three-dimensional Euclidean subalgebra)  from a particular  gauge condition which fixed the reparametrization freedom.
It should also be possible to obtain the supersymmetric Snyder algebra starting from an action principle for supersymmetric particles, for example \cite{Galvao:1980cu}.  As the system has two first class constraints it will require to gauge constraints to eliminate all gauge degrees of freedom.  There may exist a choice of conditions whereby the supersymmetric Snyder algebra is realized by the Dirac brackets.

Finally, a number of generalizations of our construction are possible.  One can consider supersymmetric Snyder algebras in different space-time dimensions and also  extended supersymmetry.    For the three-dimensional Euclidean version of the supersymmetry Snyder algebra, see \cite{Beppefest}.  Concerning  extended supersymmetry, its construction is straightforward.  For this we can introduce $2N$ de Sitter spinors $\Theta^{\tt a}_{\tt A}$ and $\bar\Theta^{\tt a}_{\tt A}$,  ${\tt a}=1,...N$, satisfying 
\beqa \{\Theta^{\tt a}_{\tt A},\bar\Theta^{\tt b}_{\tt B}\}&=&-i\delta_{\tt AB} \delta^{\tt ab}\cr &&\cr \{\bar\Theta^{\tt a}_{\tt A},\bar\Theta^{\tt b}_{\tt B}\}&=&\frac 2\Lambda E^+_{\tt AB} Z_+^{\tt ab}\cr &&\cr \{\Theta^{\tt a}_{\tt A},\Theta^{\tt b}_{\tt B}\}&=&\frac 2\Lambda E^-_{\tt AB} Z_-^{\tt ab}\;,
 \label{xtnddCapthtbrkts} \eeqa
where $Z_+^{\tt ab}=-Z_+^{\tt ba}$ and  $Z_-^{\tt ab}=-Z_-^{\tt ba}$ are central charges and we define
\be E^+=\pmatrix{0&0\cr 0&-\epsilon_{\dot\alpha\dot\beta}}
\qquad\qquad E^-=
\pmatrix{\epsilon_{\alpha\beta}&0\cr0&0}\ee
The de Sitter generators (\ref{dstrgnrtrs}) now contain a sum over  ${\tt a}=1,...N$.  Upon writing
\be \bar \Theta^{\tt a}=\frac 1{\sqrt{\Lambda}}\pmatrix{\Lambda\theta^{\tt a} \;&\;\bar Q^{\tt a} +i\theta^{\tt a}\sigma^n p_n }\qquad\qquad \Theta^{\tt a}=\frac 1{\sqrt{\Lambda}}\pmatrix{-Q^{\tt a}+ip_n\sigma^n\bar\theta^{\tt a}\cr\Lambda\bar\theta^{\tt a} }\;,\ee and substituting into (\ref{xtnddCapthtbrkts}),
we recover the extended supersymmetry algebra
\beqa
\{Q^{\tt a}_\alpha,\bar Q^{\tt b}_{\dot \alpha}\}&=& 2{\sigma_{\alpha\dot\alpha}}^m p_m\delta^{\tt ab}\cr &&\cr\{Q^{\tt a}_\alpha,Q^{\tt b}_\beta\}&=&2\epsilon_{\alpha\beta}Z_-^{\tt ab}
\cr &&\cr\{\bar Q^{\tt a}_{\dot \alpha},\bar Q^{\tt b}_{\dot \beta}\}&=&-2\epsilon_{\dot\alpha\dot\beta}Z_+^{\tt ab}\;,
\eeqa
upon projecting to Minkowski space-time.   From (\ref{xtnddCapthtbrkts}), it also follows that 
$\theta^{{\tt a}\alpha}$ and $\bar\theta^{{\tt a}\dot\alpha}$ are Grassmann odd variables, which are canonically conjugate, respectively, to $Q^{\tt a}_\alpha$ and $\bar Q^{\tt a}_{\dot\alpha}$, i.e.,
  \be
 \{Q^{\tt a}_\alpha,\theta^{\beta^{\tt b}}\}=i\delta_{\alpha}^{\beta}\delta^{\tt ab}\qquad \quad \{\bar Q^{\tt a}_{\dot \alpha},\bar\theta^{\dot\beta{\tt b}}\}=-i\delta_{\dot \alpha}^{\dot \beta}\delta^{\tt ab}\qquad\quad \{Q^{\tt a}_\alpha,\bar\theta^{\dot\beta{\tt b}}\}= \{\bar Q^{\tt a}_{\dot \alpha},\theta^{\beta{\tt b}}\}=0 \label{susyhisnbrg}\;\ee

\bigskip

{\Large {\bf Acknowledgments} }

\noindent
L. G. was  supported by the High Energy Section of ICTP.
A.S. was supported in part by the DOE,
Grant No. DE-FG02-10ER41714.

\end{document}